# Alpha Decay of Superheavy Nuclei

*Frank Bello, Javier Aguilera, Oscar Rodríguez*

InSTEC, La Habana, Cuba

frankl@instec.cu

## Abstract

Recently synthesis of superheavy nuclei has been achieved in hot fusion reactions. A systematic theoretical calculation of alpha decay half-lives in this region of the periodic system, may be useful in the identification of new nuclei in these type of reactions. Alpha decay half-lives are obtained in the framework of the barrier penetration theory, built with the use of proximity and Coulomb forces. It is estimated from a classical viewpoint, a possible angular momentum maximum value for odd and odd-odd nuclei alpha particles. The masses and deformations of nuclei are obtained from a macro-microscopic model, with the use of the two center shell model. Decay half-lives are compared with experimental results.

## 1. Introduction

One of the main problems of modern nuclear physics is the extension of the periodic system into the islands of stability of superheavy elements (SHE). For the synthesis of these nuclei fusion-evaporation reactions are used and two approaches have been successfully employed: cold and hot fusion. The former have been used to produce new elements and isotopes up to $Z=113$ [1,2], and the last one have been used to produce more neutron rich isotopes of elements up to $Z=118$ [3]. The identification of SHEs in cold fusion reactions is based on the identification of the products via alpha correlations with known alpha emitters at the end of the decay sequences, but in hot fusion reactions the nuclei at the end of the decay sequences are neutron rich isotopes that have not been obtained yet in other kind of experiments; thus in this type of reactions an alpha-decay systematics based on theoretical calculations provide a useful tool for an ulterior identification of the reaction products.

## 2. Half-live Calculation.

In the quantum tunneling theory of alpha decay, the decay constant $\lambda$ can be expressed as the product of the alpha particle preformation probability $P_0$, by the number of assaults on the barrier per second $\nu$, by the barrier penetration probability $P$.

$$\lambda = P_0 \nu P \qquad (1)$$

The barrier penetration probability is calculated using one-dimensional WKB approximation

$$P = \exp\{-\frac{2}{\hbar}\int_{z_1}^{z_2} \sqrt{2\mu(V(z)-Q_\alpha)}\,dz\} \qquad (2)$$

where $m$ is the reduced mass. The potential energy $V$ is the sum of the Coulomb $V_C$, nuclear $V_N$ and centrifugal $V_l$ energy.

$$V(z) = V_C(r) + V_N(z) + V_l(r) \qquad (3)$$

In the above expressions, $z$ and $r$ are, respectively, the distances between the surfaces and between the centers of the alpha particle and the residual nucleus, both measured along an axis parallel to the vector describing the relative motion; in the present work is considered that the alpha particle is emitted from the farthest point of the nuclear surface (see Figure 1), because in this way the alpha particle face a lesser Coulomb barrier. The turning points $z_1$ and $z_2$ are determined from the equation

$$V(z_1) = V(z_2) = Q_\alpha \qquad (4)$$

For $P_0$ and $\nu$, values in [4] have been taken; they have been obtained by a fit with a selected set of experimental data. For even-even, even-odd, odd-even and odd-odd nuclei we have

$$\begin{aligned} c_{ee} &= -20.198 & c_{eo} &= -19.412 \\ c_{oe} &= -19.680 & c_{oo} &= -18.903 \end{aligned} \qquad (5)$$

where

$$c = \log_{10}(\ln(2)) - \log_{10}(P_0) - \log_{10}(\nu) \qquad (6)$$

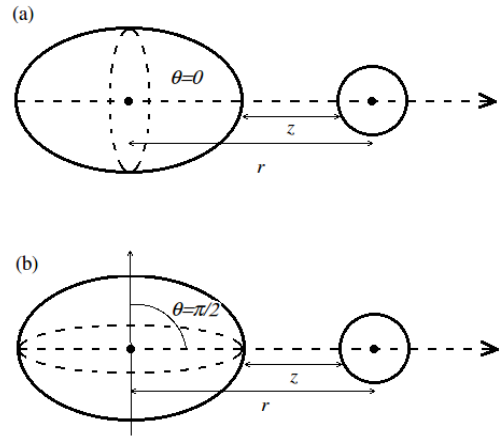

Figure 1: *The alpha particle is emitted from the farthest point of the nuclear surface.* (a) *case of prolate nucleus.* (b) *case of oblate nucleus.*

### 2.1. Proximity Potential.

The nucleus is a leptodermic distribution, i. e., a distribution essentially homogeneous except for its surface. The strong attraction between two nuclei occurs when their surfaces approach to a distance comparable to the surface width $b$; the energy of this interaction can be described by the proximity potential [5].

$$V_N(z) = 4\pi\gamma b \bar{K} \Phi(z/b) \qquad (7)$$

Here $\gamma$ is the nuclear surface tension coefficient, $\bar{K}$ is the reciprocal of the square root of the Gaussian curvature of the function that determines the distance between two points of the surfaces, evaluated at the point of closest approach, and $\Phi$ is an adimensional function called universal proximity potential.

For the surface tension coefficient is used the Lysekil formula [5]

$$\gamma = \gamma_0(1 - 1.7826\,I^2) \qquad (8)$$

in which $\gamma_0 \approx 0.95\,MeV/fm^{-2}$ and $I = (N-Z)/A$, where $N$, $Z$ and $A$ refer to the set of both nuclei.

Calculation of the Gaussian curvature of a function that depends on the surfaces shapes of two deformed nuclei can be difficult; $\bar{K}$ in (7) can be replaced by a simple expression that depends on the principal curvatures $k_i^x$, $k_i^y$ of the surfaces of both nuclei.

$$\bar{K}' = [(k_1^x + k_2^x)(k_1^y + k_2^y)]^{1/2} \quad (9)$$

In the development in spherical harmonics $Y_l^m$ of the nucleus surface, generally is only taken into account quadrupole deformations; therefore, for a nucleus with axial symmetry, the radius $R_N$ can be expressed, depending on the parameter $\beta_2$, as follows:

$$R_N(\theta) = C(1 + \beta_2 Y_2^0(\theta)) \quad (10)$$

Here $C$ is the radius of a spherical nucleus with the same volume as the deformed nucleus. To define the radius of a leptodermic distribution there are several parameters, the best known is the sharp radius, usually taken as $R = R_0 A^{1/3}$. However, when the proximity potential is used, it is preferable to take the nucleus radius as the central radius [6], which is determined mostly by the characteristics of the nucleus surface and not by the value of the density distribution function inside the nucleus. The central radius is related with the sharp radius by the expression:

$$C = R - \frac{b^2}{R} \quad (11)$$

The next formula [5] can be used for the sharp radius

$$R = 1.28 A^{1/3} - 0.76 + 0.8 A^{-1/3} \quad (12)$$

which take into account an $R_0$ dependence with $A$.

From (10), the principal curvatures of the nucleus in the alpha particle emission point are calculated; for prolate nuclei (Figure 1 (a))

$$k_N^x = k_N^y = \frac{1 + 5B + 4B^2}{C(1+B)^3} \quad (13)$$

and for oblate nuclei (Figure 1 (b))

$$k_N^x = \frac{1 - 4B + (7/4)B^2}{C(1-(1/2)B)^3}, \quad k_N^y = \frac{1}{C(1-(1/2)B)} \quad (14)$$

$$\text{where} \quad B = \sqrt{\frac{5}{4\pi}} \beta_2 \quad (15)$$

Alpha particle curvature is equal to the inverse of its radius, in this case the central radius; we take its sharp radius as $R_\alpha = 1.671\, fm$. The universal proximity potential [7] was obtained from the Fermi gas model with the inclusion of a momentum dependent nucleon-nucleon interaction potential; it reads:

$$\Phi(z/b) = -1.7817 + 0.9270 \frac{z}{b} + 0.0169\left(\frac{z}{b}\right)^2 - 0.05148\left(\frac{z}{b}\right)^3$$

$$\text{for} \quad 0 \leq \frac{z}{b} \leq 1.9475, \quad (16\,a)$$

and

$$\Phi(z/b) = -4.41 e^{-\frac{z}{0.7176b}} \quad \text{for} \quad 1.9475 \leq z/b \quad (16\,b)$$

## 2.2. Coulomb Energy.

For two nuclei axially symmetric with quadrupole deformations, the Coulomb interaction can be expressed analytically [8]; in the case of the nucleus and the alpha particle we have

$$V_c(r,\theta) = 2Ze^2[F^{(0)}(r) + F_2^{(1)}(r)Y_2^0(\theta)\beta_2 + (\frac{\sqrt{5}}{7}\frac{1}{\sqrt{\pi}}F_2^{(2)}(r)Y_2^0(\theta) + \frac{3}{7}\frac{1}{\sqrt{\pi}}F_4^{(2)}(r)Y_4^0(\theta))\beta_2^2] \quad (17)$$

where the $F_\lambda^{(n)}(r)$ are form factors and $\theta$ is the angle between the nucleus symmetry axis and the direction of relative motion (see Figure 1). The form factors are:

$$F^{(0)}(r) = \frac{1}{r}, \quad F_2^{(1)}(r) = \frac{3}{5}\frac{C^2}{r^3},$$

$$F_2^{(2)}(r) = \frac{6}{5}\frac{C^2}{r^3}, \quad F_4^{(2)}(r) = \frac{C^4}{r^5} \quad (18)$$

## 2.3. Centrifugal Potential.

The orbital quantum number $l$ of the emitted alpha particle is the fundamental factor in determining the centrifugal barrier

$$V_l(r) = \frac{\hbar^2 l(l+1)}{2\mu r^2} \quad (19)$$

In this paper is considered that both the parent nucleus and the residual nucleus are in the ground state, therefore, for even-even nuclei $l=0$. For odd and odd-odd nuclei, from the classical definition of angular momentum, we can make an argument that leads to estimate a maximum value for $l$ ($l_{max}$), whose fundamental idea is that the impact parameter of the alpha particle can not be greater than the radius of the emitter nucleus; from here we obtain

$$\frac{\hbar^2 l(l+1)}{2\mu} \leq Q_\alpha R_N^2 \quad (20)$$

## 2.4. Masses and Deformations in the Ground State.

Nuclei masses were calculated by a macro-microscopic method, obtaining the interaction potential energy of two nuclei in a fusion-fission process [9]. The macroscopic part of the calculation is performed with a liquid drop model version that takes into account the finite range of the nuclear forces [10, 11], and the for the microscopic calculation, based on the shell correction method of Strutinsky [12], we use the two center shell model (TCSM) [13]. The energy of the system calculated by this method depends on five parameters; fixing three of them ($\varepsilon = 1$, $\eta = 0$ y $\delta_1 = \delta_2 = \delta$) (see [9]) we can construct a potential surface whose minimum point corresponds to the ground state. The deformation is obtained once found the minimum energy state and known the shape of the nucleus. From (12)

$$\beta_2 = \sqrt{4\pi} \frac{\int_0^\pi R_N(\theta) Y_2^0(\theta) sen\theta d\theta}{\int_0^\pi R_N(\theta) Y_0^0(\theta) sen\theta d\theta} \quad (21)$$

The shell correction and deformation values calculated for all nuclei in the $Z = 100 \rightarrow 130$, $N = 150 \rightarrow 205$ region are shown in Figures 2 and 3. In both maps there are some areas that contrast with other nearby areas showing well-defined limits, which is an unnatural behavior; as a palpable deficiency in the method to find the masses and deformations of nuclei, there is a probability that in the search for a global minimum, a local minimum be found which distorts the results. Nuclei obtained recently in hot fusion reactions at JINR [3] are indicated in both maps, observing (Figure 2) that they are in an area of relative stability; in the east-northeast is seen a wide zone of stability.

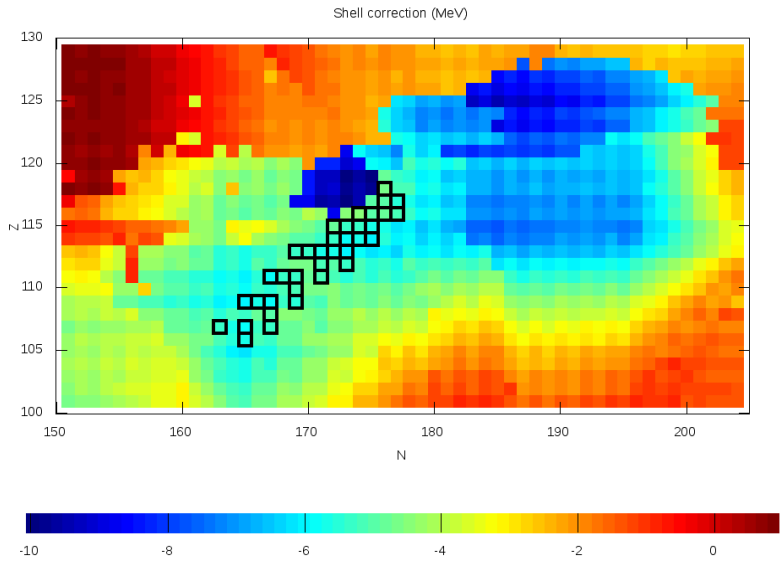

Figure 2: *Shell correction values calculated by a macro-microscopic method. Nuclei recently synthesized in hot fusion reactions are marked.*

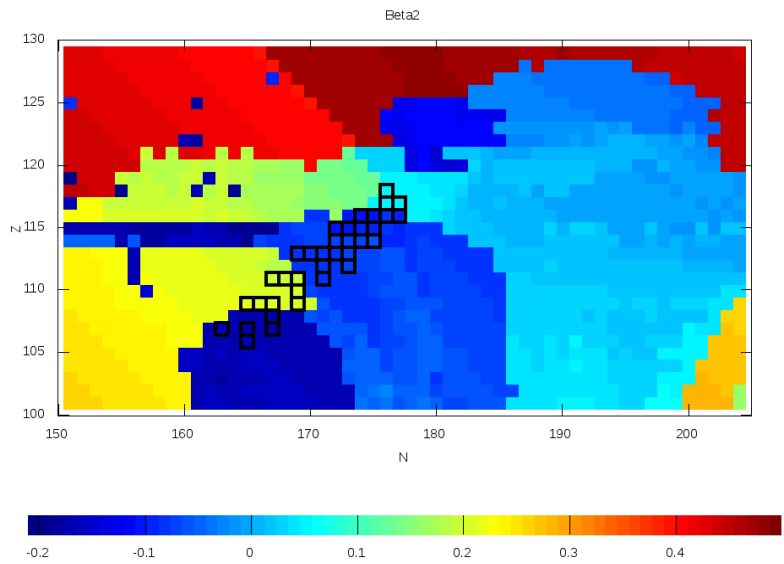

Figure 3: *Deformations values calculated by a macro-microscopic method. Nuclei recently synthesized in hot fusion reactions are marked.*

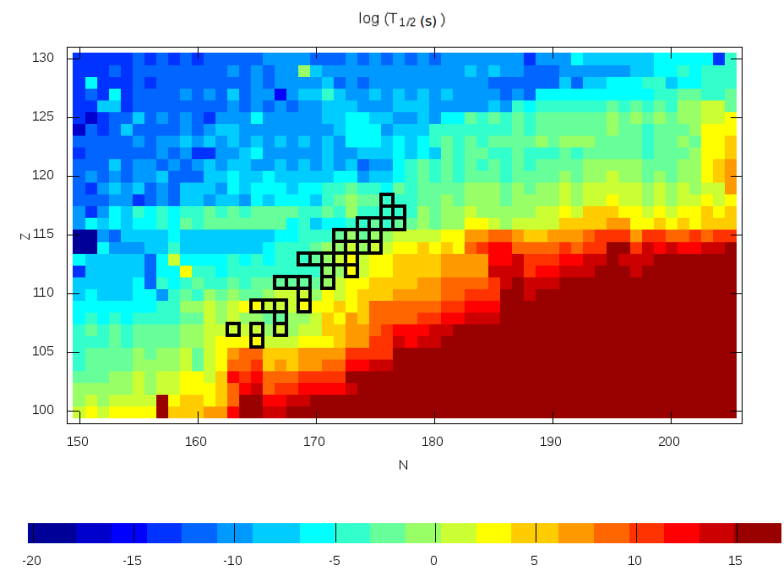

Figure 4: *Decimal logarithm of calculated alpha half-lives. For odd and odd-odd nuclei l = 0. Nuclei recently synthesized in hot fusion reactions are marked.*

## 3. Results and Discussion

Figure 4 shows the decimal logarithm of calculated alpha half-lives ($T_{1/2}$) for all nuclei in the region $Z=100\rightarrow130$, $N=150\rightarrow205$; for odd and odd-odd nuclei it shows the values obtained with $l = 0$. The figure also displayed the nuclei synthesized in hot fusion reactions. The map distinguishes three zones: one blue and one brown area corresponding to extremely unstable and very stable nuclei, respectively, relative to alpha decay, and the green-yellow zone corresponding to those for which a possible alpha decay occur with a half-life with an order from $10^{-3}$ to $10^3$ s or so. Some values shown in Figure 4 are tabulated in the Annex.

### 3.1. Even-even Nuclei.

Figures 5 and 6 compare calculated half-lives with experimental values (blue) obtained at JIRN, of the chains of nuclei $116^{292}$ and $118^{294}$ [3]. Firstly Figures 5 (a) and 6 (a) show the results obtained from the Viola-Seaborg formula (VSS) [14] (red) and the formalism of the barrier penetration theory (BPT) (black), as in Figures 5 (b) and 6 (b), but in this case we did not include the deformations of nuclei in the BPT calculation. Parameters of the VSS formula were taken from [15]. It can be seen that $T_{1/2}$ values for the nuclei $116^{292}$ and $118^{294}$ worsens significantly when we include deformations in calculation, this coincides with an abrupt jump in deformations values next to these nuclei on Figure 3.

Figure 7 shows the results of BPT calculation made with the masses and the deformations calculated using the TCSM (black), and also with the masses and deformations reported by Möller [11] (red). The two isotopes of element 114 shows a good correspondence in the case of TCSM, but not for Möller; however, for the two isotopes of element 116 and for $118^{294}$, Möller calculations are better. In map 2 can be seen that nuclei $116^{290}$ and $118^{294}$ occupy a position immediately next to an area where there is a sharp increase of stability; the even isotopes of the element 114, which yield the best results, are found in both maps (Figures 2 and 3) in areas more distant from the frontiers.

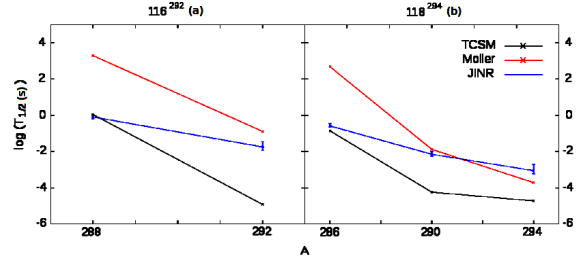

Figure 7: *BPT calculations with TCSM masses and deformations and Möller masses and deformations for the $116^{292}$ (a) and $118^{294}$ (b) chains. Experimental values (JINR) are included.*

### 3.2. Odd and Odd-odd Nuclei.

Figures 8-14 show the results obtained for the chains of nuclei $113^{282}$, $115^{287}$, $115^{288}$, $116^{291}$, $116^{293}$, $117^{293}$ and $117^{294}$; experimental values are in blue. BPT calculations (black) in Figures 8 (a) -14 (a) were performed with $l = 0$, but Figures 8 (b) -14 (b) shows a range of $T_{1/2}$ values given by the variation of the angular momentum of alpha particle from zero to $l_{max}$ (see **2.3**); also appear in these figures the results of the VSS formula (red).

In general, BPT calculation for $l = 0$ differs from VSS formula more widely than in the case of even-even nuclei, because in general, $l$ has nonzero value. As can be seen in Figures 8 (a) -14 (a), the value given by VSS is included in the range determined by the variation of $l$ except in some cases, for example, the nuclei $115^{287}$, $115^{288}$ and $116^{293}$ whose deformations are at border areas; in the case of some nuclei such as $116^{291}$ and other from chains of $116^{293}$ and $117^{293}$, the difference of VSS with the value of BPT for $l = 0$ is small. The previous results are in good enough agreement with experiment, taking into account the margin of error that causes the angular momentum variation, with the exception of a few, for example, some isotopes of Meitnerium ($A = 109$).

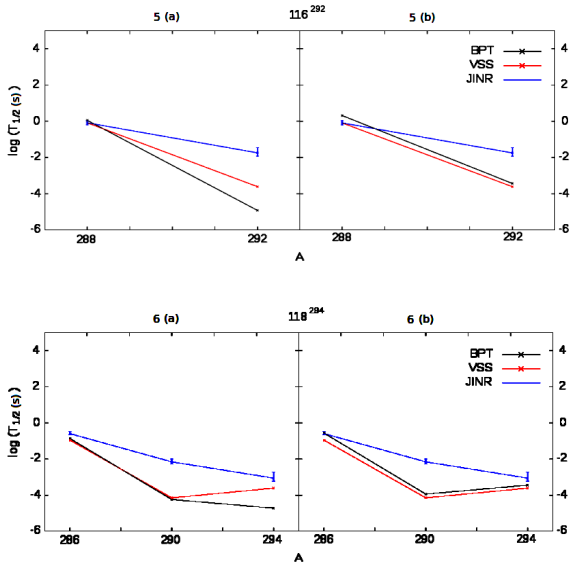

Figures 5 and 6: *Comparison of VSS and BPT calculations with experiment (JINR), for $116^{292}$ (5) and $118^{294}$ (6) chains. In (b), deformations are excluded of calculations.*

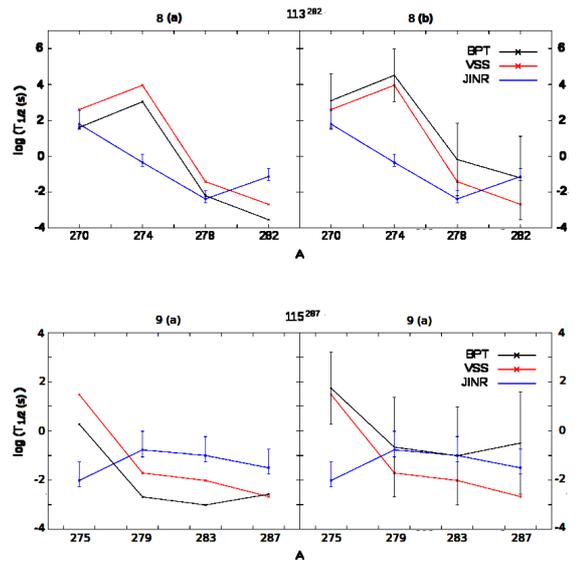

Figures 8 and 9: *VSS and BPT calculations and experiment values for $113^{282}$ and $115^{287}$ chains. In (a) $l=0$; in (b) $l=0$ to $l_{max}$, and the midpoints of the intervals are jointed.*

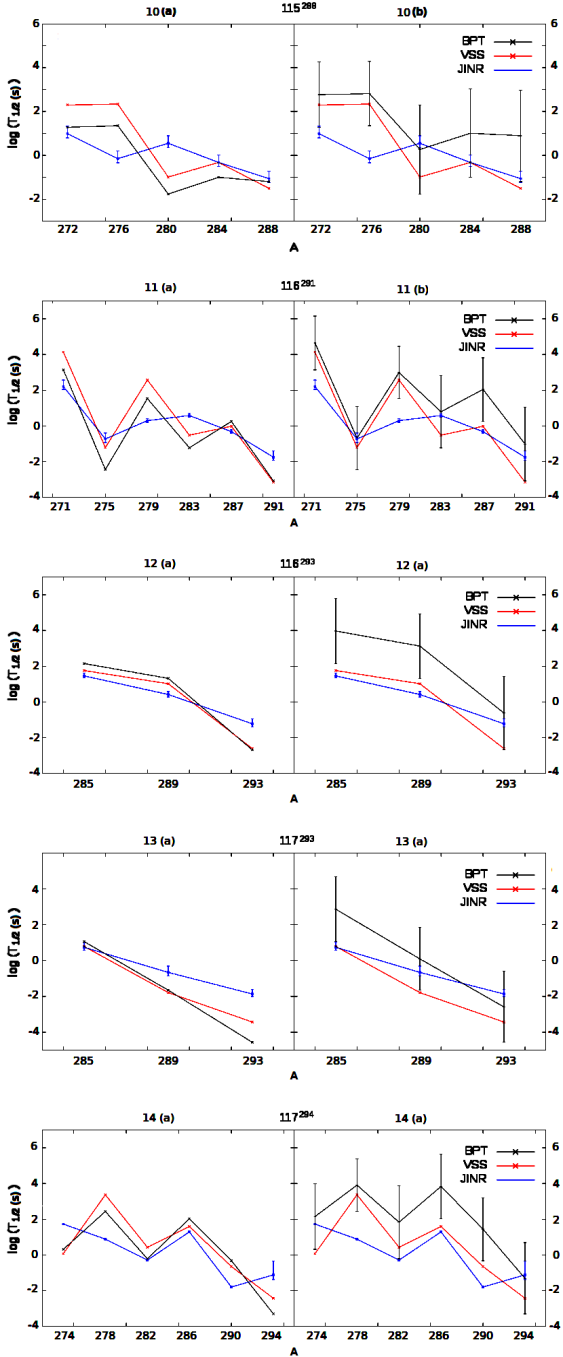

Figures 10 to 14: *Comparison of VSS and BPT calculations with experiment (JINR) for* $115^{288}$, $116^{291}$, $116^{293}$, $117^{293}$ *and* $117^{294}$. *In* (a) *l=0; in (b) l=0 to* $l_{max}$, *and the midpoints of the intervals are attached.*

In Figure 15, as in Figure 7, we compare the results with the masses and deformations of Möller.

Taking into account all nuclei of the chains mentioned so far (including even-even nuclei) we can take the standard deviation $\sigma$ of the calculation with respect to the experiment as a way of comparison

$$\sigma = \left[ \frac{1}{N-1} \sum_{i=1}^{N} (log_{10} T_{1/2\,i} - log_{10} T_{1/2\,i\,exp})^2 \right]^{1/2} \quad (22)$$

In Möller case $\sigma = 2.51$ ($\sigma = 0.82$ for even-even nuclei) and in TCSM case $\sigma = 1.61$ ($\sigma = 0.70$ for even-even nuclei). In all cases we take values for $l = 0$.

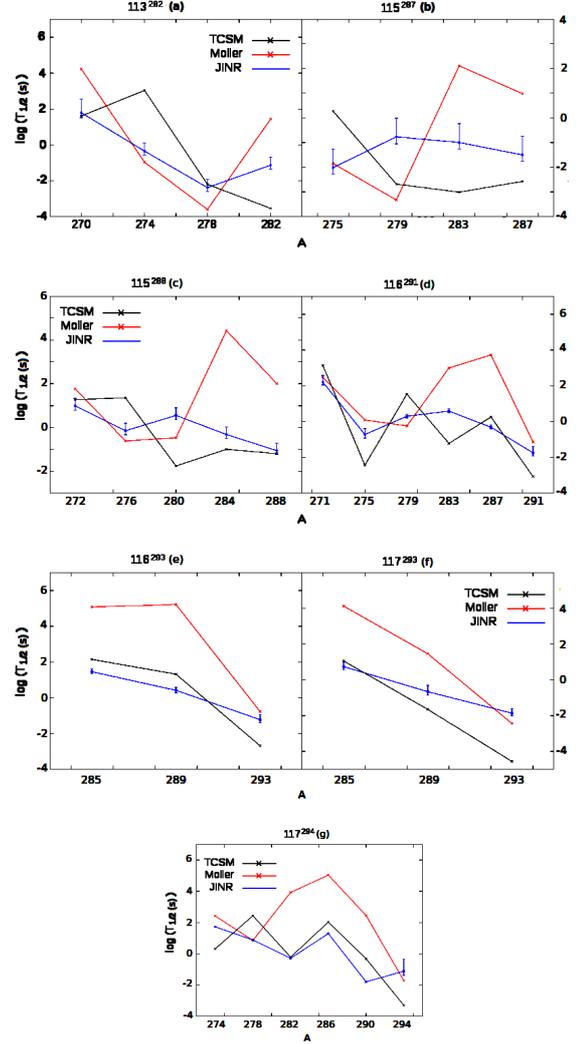

Figure 15: *BPT calculations with TCSM masses and deformations and Möller masses and deformations for* (a) $113^{282}$, (b) $115^{287}$, (c) $115^{288}$, (d) $116^{291}$, (e) $116^{293}$, (f) $117^{293}$ *and* (g) $117^{294}$ *chains. Experimental values (JINR) are included.*

## 4. Acknowledgements

Authors wants to thank Yaser Martinez and Luis Felipe from Prof. W Greiner work group at FIAS for the ground states masses and deformations calculated using the two center shell model.

## Annex

Table 1 shows the decimal logarithm of calculated alpha half-lives ($T_{1/2}$) for some nuclei in the region $Z = 100 \rightarrow 130$, $N = 150 \rightarrow 205$; for odd and odd-odd nuclei it shows the values obtained with $l = 0$. T stand for $\log_{10}(T_{1/2}(s))$.

| Z=100 | | Z=101 | | Z=102 | | Z=103 | |
|---|---|---|---|---|---|---|---|
| A | T | A | T | A | T | A | T |
| 250 | 1.67 | 251 | 0.10 | 252 | -1.42 | 253 | -2.15 |
| 251 | 2.08 | 252 | 0.68 | 253 | -0.78 | 254 | -1.65 |
| 252 | 1.82 | 253 | 0.21 | 254 | -1.27 | 255 | -2.07 |
| 253 | 3.03 | 254 | 1.28 | 255 | -0.14 | 256 | -0.73 |
| 254 | 2.83 | 255 | 0.85 | 256 | -0.68 | 257 | -1.16 |
| 255 | 3.71 | 256 | 1.86 | 257 | 0.58 | 258 | 0.23 |
| 256 | 3.63 | 257 | 1.51 | 258 | 0.05 | 259 | -0.12 |
| 257 | | | | | | | |
| 258 | 4.43 | 259 | 2.61 | 259 | 1.79 | 260 | 1.48 |
| | | 260 | 4.09 | 260 | 1.59 | 261 | 1.30 |
| | | 261 | 3.84 | 261 | 3.15 | 262 | 2.68 |
| | | 262 | 2.35 | 262 | 2.90 | 263 | 2.42 |
| | | | | 263 | 2.18 | 264 | 1.89 |
| | | | | 264 | 4.72 | 265 | 4.26 |

| Z=104 | | Z=105 | | Z=106 | | Z=107 | |
|---|---|---|---|---|---|---|---|
| A | T | A | T | A | T | A | T |
| 254 | -3.74 | 255 | -3.74 | 257 | -3.55 | 258 | -3.08 |
| 255 | -2.60 | 256 | -2.65 | 258 | -3.95 | 259 | -3.79 |
| 256 | -3.12 | 257 | -3.18 | 259 | -3.18 | 260 | -3.03 |
| 257 | -1.82 | 258 | -1.90 | 260 | -3.60 | 261 | -3.45 |
| 258 | -2.24 | 259 | -2.33 | 261 | -2.55 | 262 | -2.42 |
| 259 | -0.83 | 260 | -1.29 | 262 | -3.03 | 263 | -2.91 |
| 260 | -1.15 | 261 | -1.72 | 263 | -2.08 | 264 | -1.97 |
| 261 | 0.09 | 262 | -0.68 | 264 | -2.27 | 265 | -2.16 |
| 262 | -0.02 | 263 | -0.79 | 265 | -1.42 | 266 | -1.02 |
| 263 | 0.98 | 264 | 0.23 | 266 | -1.35 | 267 | -0.99 |
| 264 | -2.05 | 265 | -1.99 | 267 | 0.89 | 268 | 1.16 |
| 265 | 0.54 | 266 | 0.50 | 268 | 2.04 | 269 | 0.29 |
| 266 | 2.79 | 267 | 2.52 | 269 | 3.53 | 270 | 1.60 |
| | | 271 | 4.92 | 270 | 2.70 | 271 | 0.92 |
| | | | | 271 | 3.14 | 272 | 1.28 |
| | | | | 272 | 1.12 | 273 | -0.41 |
| | | | | 273 | 1.74 | 274 | 0.31 |
| | | | | 274 | 1.49 | 275 | -0.01 |
| | | | | 275 | 2.09 | 276 | 0.96 |
| | | | | 276 | 2.29 | 277 | 0.69 |
| | | | | 277 | 3.32 | 278 | 4.90 |
| | | | | | | 279 | 4.60 |

| Z=108 | | Z=109 | | Z=110 | | Z=111 | |
|---|---|---|---|---|---|---|---|
| A | T | A | T | A | T | A | T |
| 263 | -4.41 | 266 | -3.50 | 269 | -2.75 | 270 | -3.97 |
| 264 | -4.99 | 267 | -3.33 | 270 | -3.32 | 271 | -3.08 |
| 265 | -4.05 | 268 | -1.41 | 271 | -0.71 | 272 | -3.40 |
| 266 | -3.87 | 269 | -1.47 | 272 | -2.82 | 273 | -3.87 |
| 267 | -1.94 | 270 | 0.45 | 273 | -1.43 | 274 | -2.62 |
| 268 | -2.00 | 271 | -0.16 | 274 | -2.40 | 275 | -3.49 |
| 269 | 0.36 | 272 | 1.26 | 275 | -1.58 | 276 | -2.67 |
| 270 | -0.24 | 273 | 2.46 | 276 | -0.63 | 277 | -3.23 |
| 271 | 1.21 | 274 | 3.03 | 277 | 0.42 | 278 | -2.20 |
| 272 | -0.86 | 275 | 0.28 | 278 | 0.31 | 279 | -2.67 |
| 273 | -0.41 | 276 | 1.35 | 279 | 1.53 | 280 | -1.75 |
| 274 | -3.15 | 277 | 1.31 | 280 | -0.05 | 281 | -2.50 |
| 275 | -2.44 | 278 | 2.43 | 281 | 3.23 | 282 | -0.21 |
| 276 | -1.47 | 279 | 2.39 | 282 | 0.91 | 283 | 0.30 |
| 277 | -0.48 | 280 | 2.66 | 283 | 2.55 | 284 | 3.15 |
| 278 | 1.94 | 281 | 4.23 | 284 | 4.97 | 285 | 3.54 |
| 279 | 4.10 | 282 | 3.47 | 286 | 4.98 | 286 | 4.72 |
| 280 | 3.59 | 283 | 4.06 | | | 287 | 4.46 |

| Z=112 | | Z=113 | | Z=114 | | Z=115 | |
|---|---|---|---|---|---|---|---|
| A | T | A | T | A | T | A | T |
| 272 | -4.57 | 276 | -4.57 | 272 | -4.71 | 284 | -4.82 |
| 273 | -4.89 | 278 | -4.50 | 281 | -4.89 | 285 | -4.59 |
| 275 | -4.10 | 280 | -4.11 | 282 | -4.57 | 286 | -3.51 |
| 276 | -4.95 | 281 | -4.94 | 283 | -3.39 | 287 | -2.57 |
| 277 | -4.07 | 282 | -3.53 | 284 | -2.54 | 288 | -1.19 |
| 278 | -4.70 | 283 | -3.01 | 285 | -0.33 | 289 | -1.65 |
| 279 | -3.58 | 284 | -1.00 | 286 | -0.85 | 290 | -0.32 |
| 280 | -4.10 | 285 | 1.06 | 287 | 0.26 | 291 | -0.81 |
| 281 | -3.96 | 286 | 2.03 | 288 | 0.03 | 292 | 0.46 |
| 282 | -3.31 | 287 | 1.60 | 289 | 1.33 | 293 | 0.23 |
| 283 | -1.23 | 288 | 3.00 | 290 | 1.22 | 294 | 1.58 |
| 284 | 0.11 | 289 | 2.76 | 291 | 2.69 | 295 | 1.30 |
| 285 | 2.15 | 290 | 4.25 | 292 | 2.44 | 296 | 2.88 |
| 286 | 1.60 | 291 | 3.81 | 293 | 3.96 | 297 | 3.48 |
| 287 | 2.89 | | | 294 | 3.41 | | |
| 288 | 2.77 | | | 295 | 4.12 | | |
| 289 | 4.30 | | | 296 | 3.65 | | |
| 290 | 3.92 | | | | | | |

| Z=116 | | Z=117 | | Z=118 | | Z=119 | |
|---|---|---|---|---|---|---|---|
| A | T | A | T | A | T | A | T |
| 273 | -3.66 | 281 | -3.35 | 290 | -2.93 | 290 | -3.28 |
| 275 | -2.79 | 282 | -2.29 | 291 | -0.83 | 291 | -4.13 |
| 276 | -3.61 | 283 | -3.17 | 292 | -2.83 | 292 | -3.05 |
| 277 | -2.29 | 284 | -1.97 | 293 | -2.10 | 293 | -3.93 |
| 278 | -3.26 | 285 | -3.14 | 294 | -4.72 | 294 | -3.73 |
| 279 | -2.19 | 286 | -3.43 | 295 | -3.44 | 296 | -3.14 |
| 280 | -3.22 | 287 | -4.73 | 296 | -3.60 | 297 | -3.73 |
| 281 | -2.21 | 288 | -2.51 | 297 | -2.11 | 298 | -2.47 |
| 282 | -2.98 | 289 | -3.37 | 298 | -2.27 | 299 | -2.61 |
| 283 | -1.59 | 290 | -1.44 | 299 | -1.34 | 300 | -1.67 |
| 285 | -4.78 | 291 | -3.55 | 300 | -1.83 | 301 | -2.11 |
| 289 | -4.94 | 292 | -4.39 | 301 | -0.35 | 302 | -1.19 |
| 290 | -4.25 | 293 | -4.55 | 302 | -0.83 | 303 | -1.26 |
| 291 | -3.06 | 294 | -3.29 | 303 | -1.16 | 304 | -1.46 |
| 292 | -4.92 | 295 | -3.49 | 304 | -1.94 | 305 | -2.61 |
| 293 | -2.67 | 296 | -1.20 | 305 | -0.72 | 306 | -1.24 |
| 294 | -2.92 | 297 | -2.02 | 306 | -1.47 | 307 | -1.78 |
| 295 | -1.38 | 298 | -0.75 | 307 | -0.01 | 308 | -0.41 |
| 296 | -2.10 | 299 | -0.99 | 308 | -0.43 | 309 | -0.85 |
| 297 | -0.67 | 300 | 0.73 | 309 | 0.44 | 310 | 0.28 |
| 298 | -0.40 | 301 | 0.22 | 310 | 0.13 | 311 | -0.19 |
| 299 | 2.35 | 302 | -0.11 | 311 | 1.01 | 312 | 0.70 |
| 300 | 2.82 | 303 | -0.70 | 312 | 0.79 | 313 | 0.26 |
| 301 | 0.42 | 304 | 0.10 | 313 | 2.18 | 314 | 1.16 |
| 302 | -0.28 | 305 | -0.30 | 314 | 1.56 | 315 | 0.77 |
| 303 | 0.74 | 306 | 0.78 | 315 | 0.61 | 316 | 0.41 |
| 304 | 0.34 | 307 | 0.51 | 316 | -0.11 | 317 | -0.59 |
| 305 | 1.46 | 308 | 3.11 | 317 | 0.95 | 318 | 0.30 |

| A | T | A | T | A | T | A | T |
|---|---|---|---|---|---|---|---|
| 306 | 1.25 | 309 | 1.53 | 318 | 0.21 | 319 | -0.40 |
| 307 | 4.09 | 310 | 4.11 | 319 | 1.23 | 320 | 0.42 |
| 308 | 3.96 | 311 | 3.97 | 320 | 0.73 | 321 | 0.20 |
| 313 | 2.92 | 313 | 3.20 | 321 | 2.10 | 322 | 1.38 |
| 314 | 2.11 | 314 | 2.17 | 322 | 1.78 | 323 | 1.45 |
| 315 | 4.79 | 315 | 1.30 | 323 | 2.95 | | |
| 316 | 2.43 | 316 | 3.71 | | | | |
| 317 | 4.69 | 317 | 1.64 | | | | |
| 318 | 3.35 | 318 | 2.62 | | | | |
| 320 | 4.92 | 319 | 2.27 | | | | |

| Z=120 | | Z=121 | | Z=122 | | Z=123 | |
|---|---|---|---|---|---|---|---|
| A | T | A | T | A | T | A | T |
| 299 | -3.09 | 299 | -3.36 | 303 | -2.89 | 303 | -3.57 |
| 300 | -3.80 | 300 | -2.42 | 304 | -3.75 | 304 | -2.77 |
| 301 | -3.26 | 301 | -3.31 | 305 | -2.73 | 305 | -3.50 |
| 302 | -4.55 | 302 | -2.86 | 306 | -3.14 | 306 | -2.59 |
| 303 | -2.35 | 303 | -3.95 | 307 | -3.82 | 307 | -3.30 |
| 304 | -2.81 | 304 | -3.00 | 308 | -4.30 | 308 | -1.74 |
| 305 | -2.89 | 305 | -3.85 | 309 | -3.26 | 309 | -2.63 |
| 306 | -4.08 | 306 | -2.49 | 310 | -4.15 | 310 | -1.88 |
| 307 | -2.58 | 307 | -3.87 | 311 | -3.28 | 311 | -2.51 |
| 308 | -2.90 | 308 | -2.84 | 312 | -3.57 | 312 | -1.10 |
| 309 | -1.85 | 309 | -3.47 | 313 | -2.81 | 313 | -1.61 |
| 310 | -2.07 | 310 | -2.56 | 314 | -3.65 | 314 | -0.76 |
| 311 | -1.28 | 311 | -3.18 | 315 | -2.89 | 315 | -1.41 |
| 312 | -1.69 | 312 | -2.37 | 316 | -2.65 | 316 | -0.43 |
| 313 | -0.23 | 313 | -1.90 | 317 | -1.54 | 317 | -2.41 |
| 314 | -0.75 | 314 | -0.76 | 318 | -2.13 | 318 | -1.86 |
| 315 | 0.64 | 315 | -1.22 | 319 | -3.14 | 319 | -2.79 |
| 316 | -0.14 | 316 | 0.18 | 320 | -3.97 | 320 | -2.53 |
| 317 | -0.88 | 317 | -0.50 | 321 | -2.50 | 321 | -3.45 |
| 318 | -2.03 | 318 | -1.31 | 322 | -3.14 | 322 | -1.69 |
| 319 | -1.06 | 319 | -2.22 | 323 | -1.57 | 323 | -2.96 |
| 320 | -1.86 | 320 | -1.89 | 324 | 0.11 | 324 | -0.62 |
| 321 | -0.53 | 321 | -1.70 | 325 | 2.41 | 325 | -1.59 |
| 322 | -0.98 | 322 | -0.85 | 326 | 3.63 | 326 | 2.61 |
| 323 | 2.79 | 323 | -0.80 | | | 327 | 3.27 |
| | | 324 | 3.12 | | | 328 | 4.72 |
| | | 325 | 3.85 | | | | |

| Z=124 | | Z=125 | | Z=126 | | Z=127 | |
|---|---|---|---|---|---|---|---|
| A | T | A | T | A | T | A | T |
| 304 | -4.57 | 308 | -3.25 | 313 | -3.44 | 326 | -3.91 |
| 305 | -4.10 | 309 | -3.87 | 315 | -4.44 | 327 | -3.69 |
| 306 | -4.66 | 310 | -2.53 | 316 | -4.89 | 328 | -1.95 |
| 307 | -3.85 | 311 | -3.31 | 317 | -4.04 | 329 | -1.77 |
| 308 | -4.58 | 312 | -2.81 | 318 | -4.58 | 330 | -4.17 |
| 309 | -3.71 | 313 | -3.37 | 319 | -3.55 | 331 | -4.10 |
| 310 | -3.43 | 314 | -2.62 | 320 | -4.16 | 332 | -2.64 |
| 311 | -2.39 | 315 | -3.07 | 321 | -3.19 | | |
| 312 | -3.90 | 316 | -2.27 | 322 | -3.91 | | |
| 313 | -3.10 | 317 | -2.81 | 323 | -3.18 | | |
| 314 | -3.04 | 318 | -1.82 | 324 | -3.77 | | |
| 315 | -2.25 | 319 | -2.38 | 325 | -3.19 | | |
| 316 | -2.72 | 320 | -1.45 | 326 | -3.67 | | |
| 317 | -1.73 | 321 | -2.03 | 327 | -1.02 | | |
| 318 | -2.28 | 322 | -1.37 | 328 | -0.79 | | |
| 319 | -1.30 | 323 | -1.96 | 329 | 1.23 | | |
| 320 | -3.20 | 324 | -1.43 | 330 | 1.11 | | |
| 321 | -2.99 | 325 | -1.90 | 331 | -2.50 | | |
| 322 | -3.94 | 326 | -1.00 | | | | |
| 323 | -2.23 | 327 | -0.90 | | | | |
| 324 | -2.80 | 328 | 1.41 | | | | |
| 325 | -1.84 | 329 | 1.72 | | | | |
| 326 | -0.04 | 330 | 3.06 | | | | |
| 327 | 2.35 | | | | | | |
| 328 | 1.99 | | | | | | |
| 329 | 3.22 | | | | | | |

Table 1. *Decimal logarithm of calculated alpha half-lives ($T_{1/2}$) for some nuclei in the region $Z=100 \rightarrow 130$, $N=150 \rightarrow 205$; for odd and odd-odd nuclei it shows the values obtained with $l = 0$. T stand for $log_{10}(T_{1/2}(s))$.*